\documentclass{svmult}

\usepackage{makeidx}         
\usepackage{graphicx}        
\usepackage{multicol}        
\usepackage{subfigure}		 
\usepackage[bottom]{footmisc}

\makeindex             

\begin{document}

\title*{Statistical Analysis of High-Flow Traffic States}
\author{Florian Knorr \and Thomas Zaksek \and
		Johannes Br\"ugmann \and Michael Schreckenberg}
\institute{Fakult\"at f\"ur Physik, Universit\"at Duisburg-Essen, 47048 Duisburg
\texttt{\{knorr,zaksek,bruegmann,schreckenberg\}@ptt.uni-due.de}}
%
%
\maketitle
\section{Introduction}
\label{sec:introduction}
\let\thefootnote\relax\footnote{JB and FK thank the German Research Foundation (DFG) 
for funding under grant no. SCHR~527/5-1.}
It is well known that traffic exhibits metastable states and hysteresis 
behavior~\cite{SchadschneiderChowdhuryNishinari2010}: 
At high vehicle flow rates, a transition from free to congested flow is 
likely to occur---resulting in a considerable decrease of the flow rate 
and significant changes of other traffic characteristics such as the 
average velocity. 
To restore high traffic flows after such a transition, it is necessary that 
the flow rate drops below a threshold value first. 

But states of high traffic flow are not only interesting from a physical 
point of view. At high flow rates, the road is operating close to its optimum. 
Therefore, it is 
is also of practical importance to investigate under what conditions 
these so-called high-flow states occur.

We will present an analysis of detector data collected from 
the motorway network of state of North Rhine-Westphalia (e.g., see 
\cite{HafsteinChrobokPottmeierSchreckenbergMazur2004,Bruegmann2013}).
This analysis focuses on the characteristics of so-called high-flow 
states (e.g., when and how often do they occur, on which lane can they 
usually be observed).
In the following, we refer to a flow rate as high, if it exceeds 
50 vehicles per minute and lane (i.e., 3000\,veh/h/lane).

\section{Analyzed Data}
\label{sec:analyzed_data}
Our analysis is based on detector data provided by more than 
3000 loop detectors during December 19, 2011 and May 31, 2013 
on the motorway network of the state of North Rhine-Westphalia (Germany).
Inductive loop detectors still are the most common source of traffic data: 
for each (1\,min)-interval, loop detectors count the number of passing vehicles, 
they measure the vehicles' velocity distinguished by vehicle type (passenger cars and trucks), 
and they determine the fraction of time within they are occupied by passing 
vehicles. 

\begin{center}
	\begin{figure}
		\subfigure[]{%
			\label{fig:high-flows-logscale}
			\includegraphics[width=0.47\textwidth]{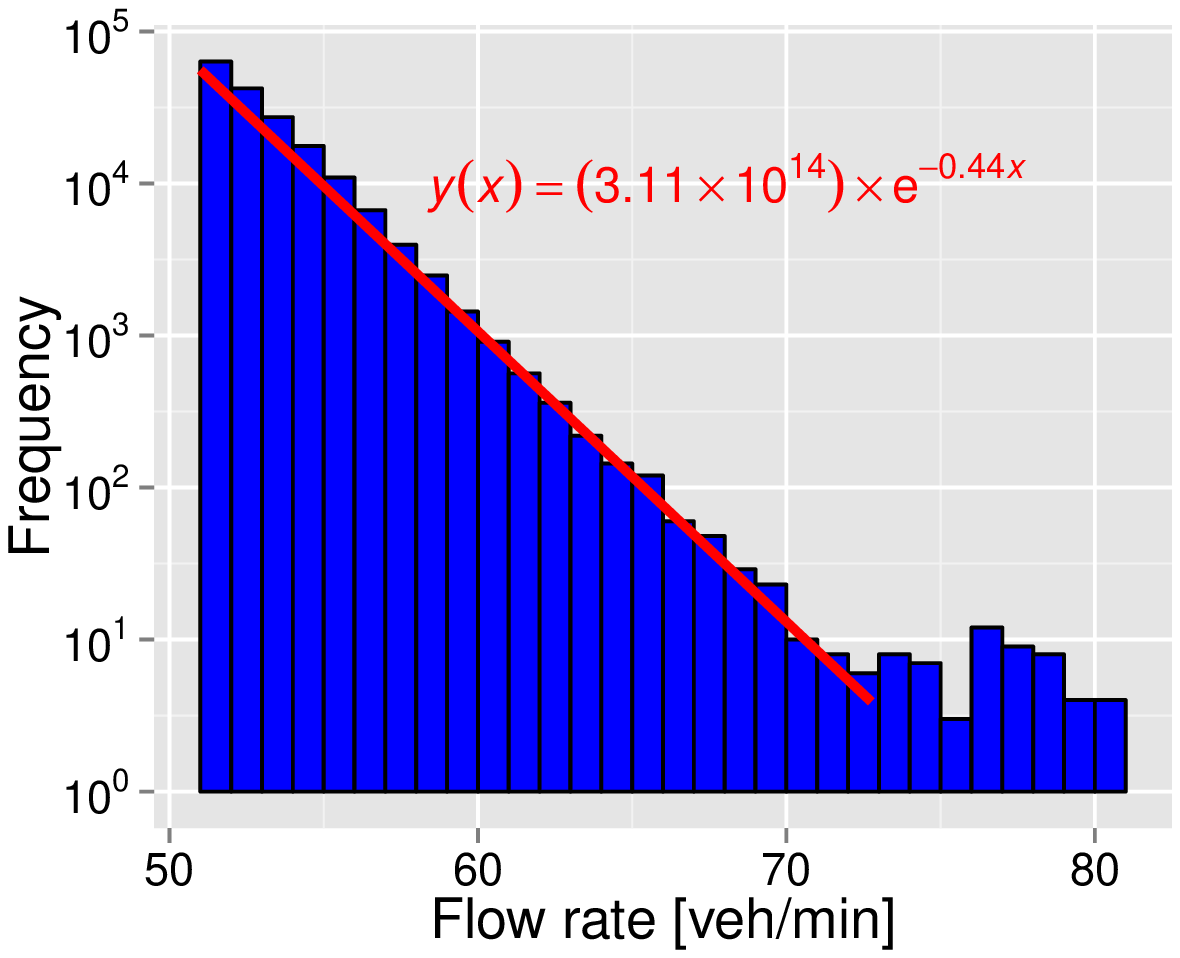}
		}
		\subfigure[]{%
			\label{fig:autobahnen}
			\includegraphics[width=0.47\textwidth]{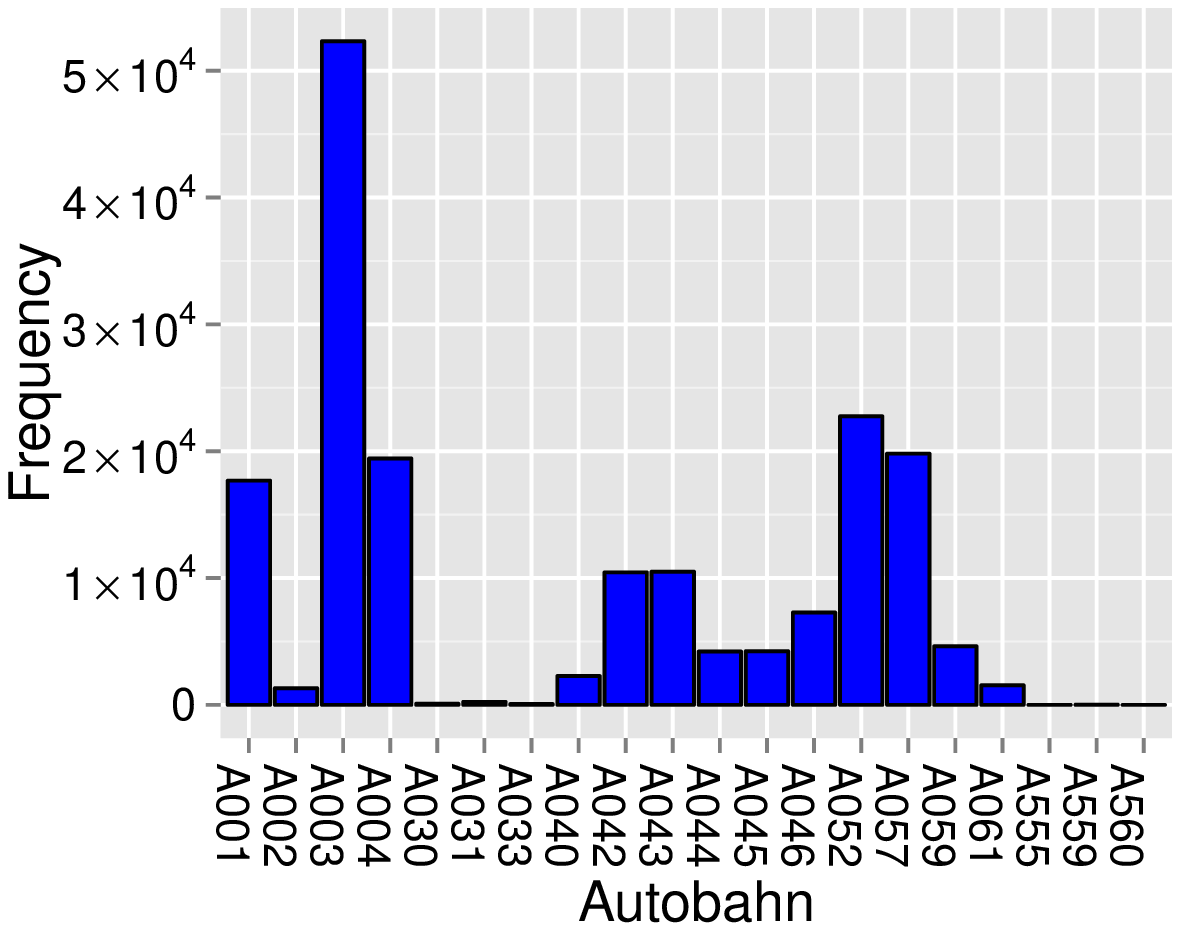}
		}
		\caption{\subref{fig:high-flows-logscale} Frequency of high-flow states  
			     and \subref{fig:autobahnen} the number of the Autobahn.}		 	
		\label{fig:autobahnen-and-flow}
 	\end{figure}
\end{center}

We  restricted our  analysis to  178\,800 measurements  exhibiting high  flow characteristics  (flow
$>$  50  veh/min) and to valid values only for each  of the just  mentioned observables.
Figure~\ref{fig:autobahnen-and-flow}  shows the  frequencies of  high-flow states  depending on  the
corresponding flow rate (\ref{fig:high-flows-logscale}).

It turns out that, up to flow rates of about 73\,veh/min, the frequencies of high-flow 
states of a given flow rate $J$ follow the power law 
$$\mathrm{frequency}=A\times\exp{\left(-\alpha\times J\right)}$$
with $A\approx 3.11\times 10^{14}$ and $\alpha\approx 0.44$.

For the small number of data sets with a higher flow rate (less than $0.03\%$, 47 total) it is not clear 
whether they actually deviate from the power law or whether these data sets indicate erroneous measurements: 
As we will see, the average velocity of high-flow states varies between 60 km/h and 120 km/h. 
Therefore, high flow rates with an average time headway of 1 s and less pose an actual risk for drivers.
Even though such time headways have already been observed at similar velocities for single vehicles \cite{AppertRoland2009}, 
it may be doubted whether this behavior can be observed for a sequence of 70 (and more) vehicles.

\section{Share of Trucks \& Lane Usage}
\begin{center}
	\begin{figure}
		\subfigure[]{%
			\label{fig:sharetrucks}
			\includegraphics[width=0.46\textwidth]{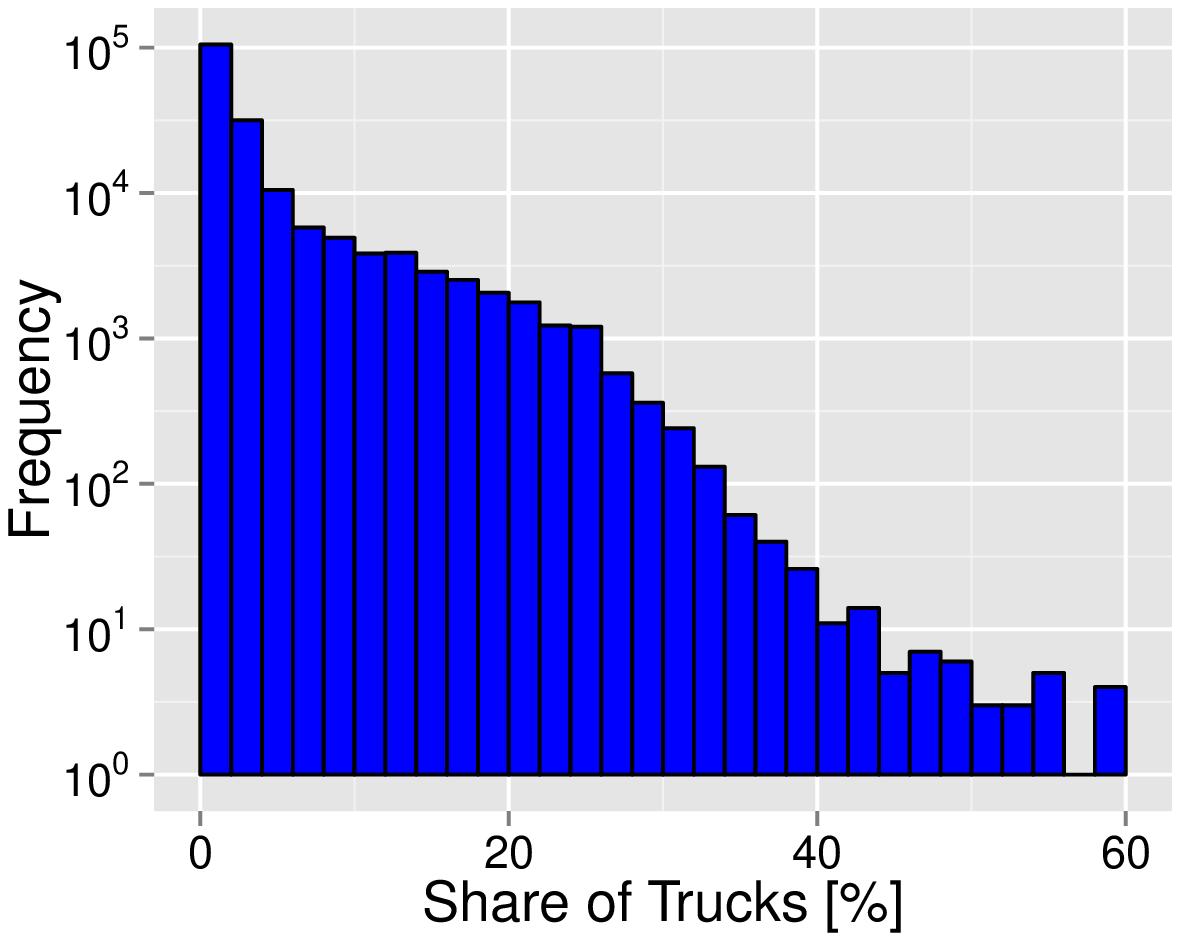}
		}
		\subfigure[]{%
			\label{fig:lanes}
			\includegraphics[width=0.46\textwidth]{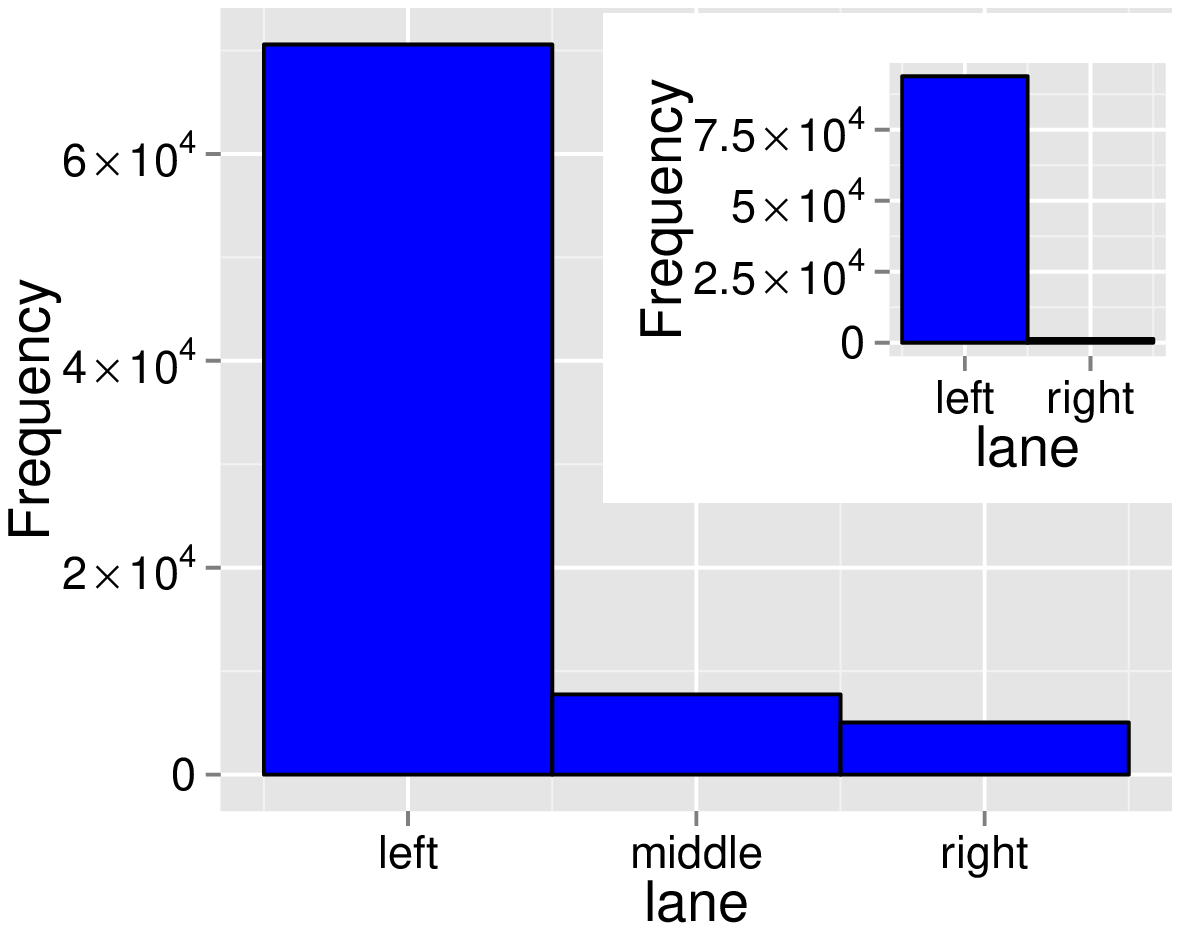}
		}
		\caption{Frequencies of high-flow traffic states 
		\subref{fig:sharetrucks} for a given share of trucks and 
		\subref{fig:lanes} resolved by the lane, on which they were observed 
		for three-lane and two-lane (inset) motorways.}
		\label{fig:trucksharelaneusage}		 			
 	\end{figure}
\end{center}

As   the   detectors  used   for   this   analysis  classify   the   detected   vehicles  into   two
groups~\cite{TLS2002},  namely  `trucks'  and  `passenger  cars',  we  can  also
investigate the  influence of  heterogeneous traffic flow  on the  occurrence of
high-flow states.  From figure~\ref{fig:sharetrucks} it becomes obvious that the
likelihood of  high traffic flows decreases  with an increasing share  of trucks
contributing to the  total flow rate. This phenomenon is  caused by trucks that
generally are restricted to a velocity of 80\,km/h in Germany. Therefore, trucks 
lower the maximum achievable average velocity in traffic flow and, due to  their 
length, they block detectors for a longer period. On the other hand, high traffic flows
require  high  average   velocities  (60\,km/h  to  120\,km/h). Consequently, 
high-flow  traffic   states  are  expected  to   favor  an  almost homogeneous 
flow of the faster passenger cars.

At  least some  high-flow traffic  measurements including  trucks may be explained  by the
classification of  the detector loops: light  trucks (e.g., SUVs and small buses / vans) are
classified as trucks, but the general speed limit for trucks does not apply to them and 
they exhibit driving characteristics similar to passenger cars.

For a similar reason the occurrence of high traffic flows is practically limited to the leftmost 
lane (see figure~\ref{fig:lanes}).
In Germany, the leftmost lane is reserved for fast-traveling vehicles to overtake, whereas 
the right lane is reserved for slow vehicles (i.e., usually trucks). 
Therefore, traffic flow on the left and middle lanes is characterized by a relatively low share of trucks and high average velocities, which facilitates the formation of high flow rates.

These results also confirm an observation first made by Sparmann~\cite{Sparmann1978} and Leutzbach and Busch~\cite{LeutzbachBusch1984} that is known a \textit{lane inversion}: 
They found that, close to the optimal flow rate, the vehicle density in the left lane surmounts the density in the right lane. 
This is surprising as by German law drivers are required to use the right (and middle) lane whenever possible. 

\section{Average Velocities of High-Flow Traffic States}
\begin{center}
	\begin{figure}
		\subfigure[]{%
			\label{fig:avg-v}
			\includegraphics[width=0.46\textwidth]{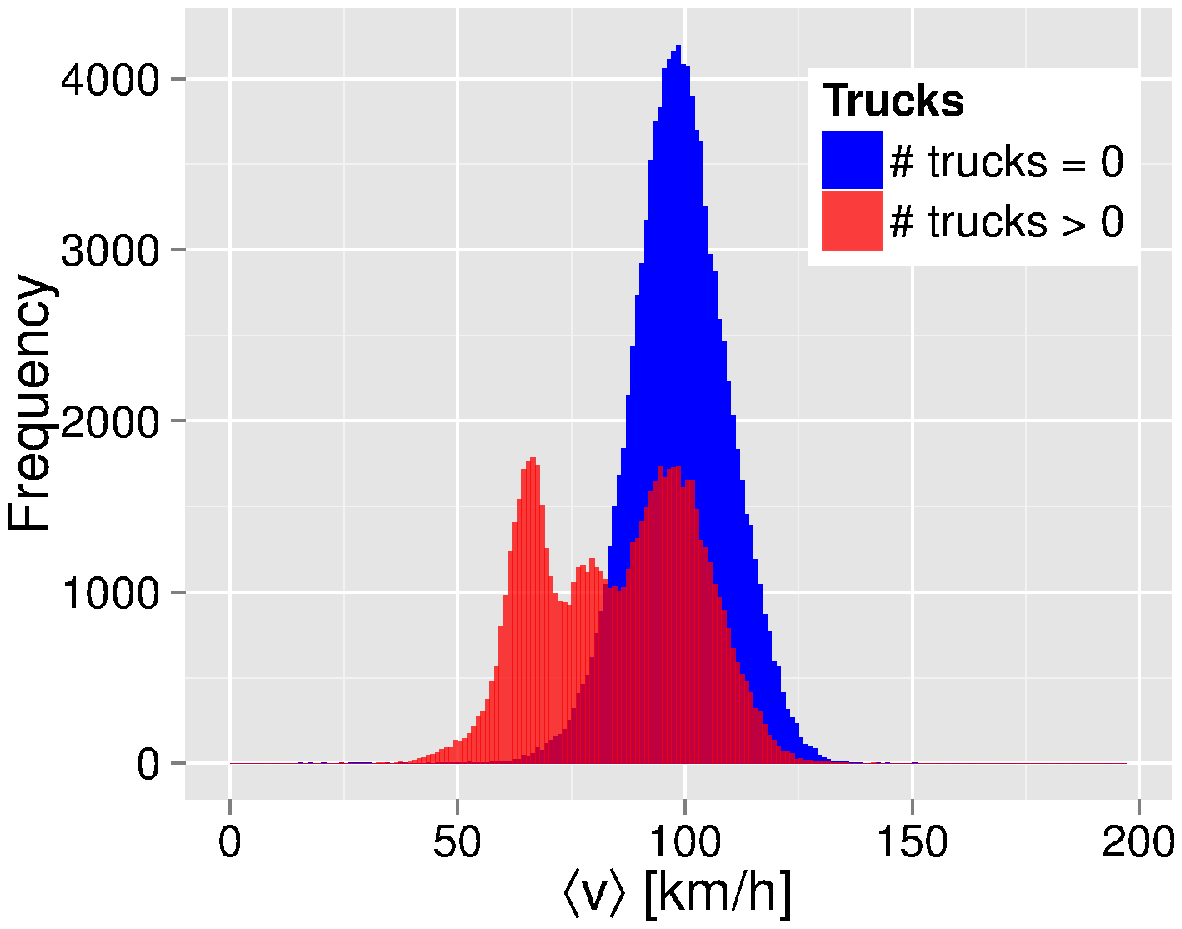}
		}
		\subfigure[]{%
			\label{fig:vel-diff}
			\includegraphics[width=0.46\textwidth]{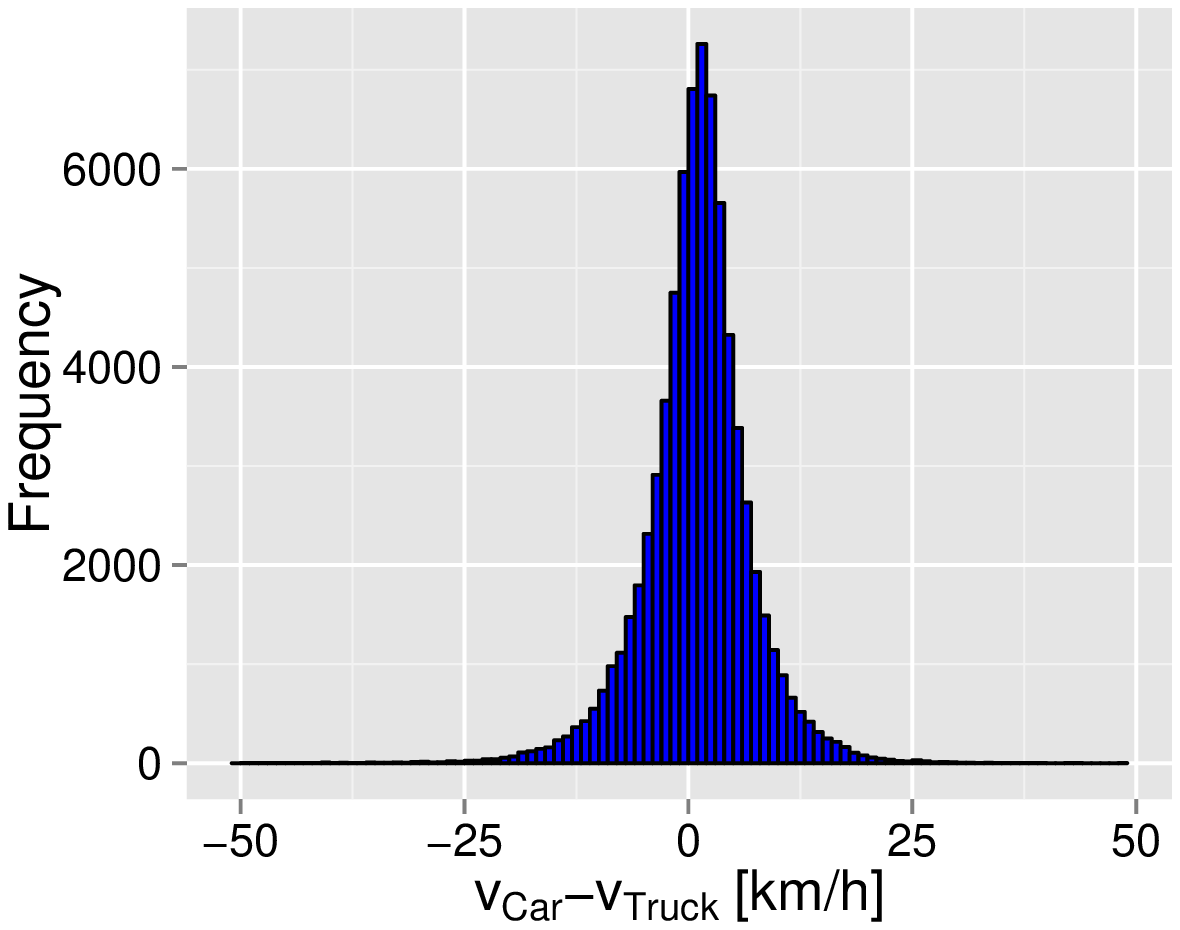}
		}
		\caption{Average velocities $\langle v \rangle$ of high-flow states and 
				the difference 
				in the average velocities of trucks ($v_\mathrm{Truck}$) 
				and cars ($v_\mathrm{Car}$).}
		\label{fig:velocities}		 			
 	\end{figure}
\end{center}

As already stated in the previous section, the average velocities of high-flow states range from approximately 60\,km/h to 120\,km/h. 
In figure \ref{fig:avg-v}, one can see the distribution of the average velocities. 
For better analysis, these measurements were subdivided into two classes: 
(i) measurements without trucks (in this case the depicted average velocity is identical to 
	the average velocity of all cars on the road) and 
(ii) measurements in which at least one vehicle was identified as a truck.

From these histograms we see that the average velocity in homogeneous traffic, 
consisting of passenger cars only, is considerably higher than in mixed traffic with 
passenger cars and trucks with a single peak at approximately 100\,km/h.
In mixed traffic, we observe two peaks: one at 100\,km/h and another one at approximately 
at 60\,km/h. 
The first one (at 100\,km/h) corresponds to a very low number of trucks contributing to the total flow. The second one results from measurements with roughly four or more trucks. 
This could be verified by varying the threshold (i.e., the number of trucks) which separate the two curves [not depicted]. 
It should also be noted that the classification of vehicle types, which is mostly based on the 
estimated vehicle length~\cite{TLS2002}, is not free of fault. 
Therefore, it is safe to assume that a certain amount of the measurements contributing especially to the first peak consisted of passenger cars only.

Figure \ref{fig:vel-diff} shows the difference in the average velocities of trucks and passenger 
cars for the observed high flow states. 
This histogram illustrates very well the synchronization of average velocities in high traffic flow: the distribution's mean is at 0\,km/h ($0.52$\,km/h) with a variance of $\sigma^2\approx 35\,\left(\mathrm{km/h}\right)^2$. 
Moreover, the resulting distribution is strongly peaked around its mean (leptokuric with a sample excess kurtosis of $3.89$).

\section{Temporal Occurrences \& Lifetimes of High-Flow Traffic States} %
\begin{center}
	\begin{figure}
		\subfigure[]{%
			\label{fig:month}
			\includegraphics[width=0.46\textwidth]{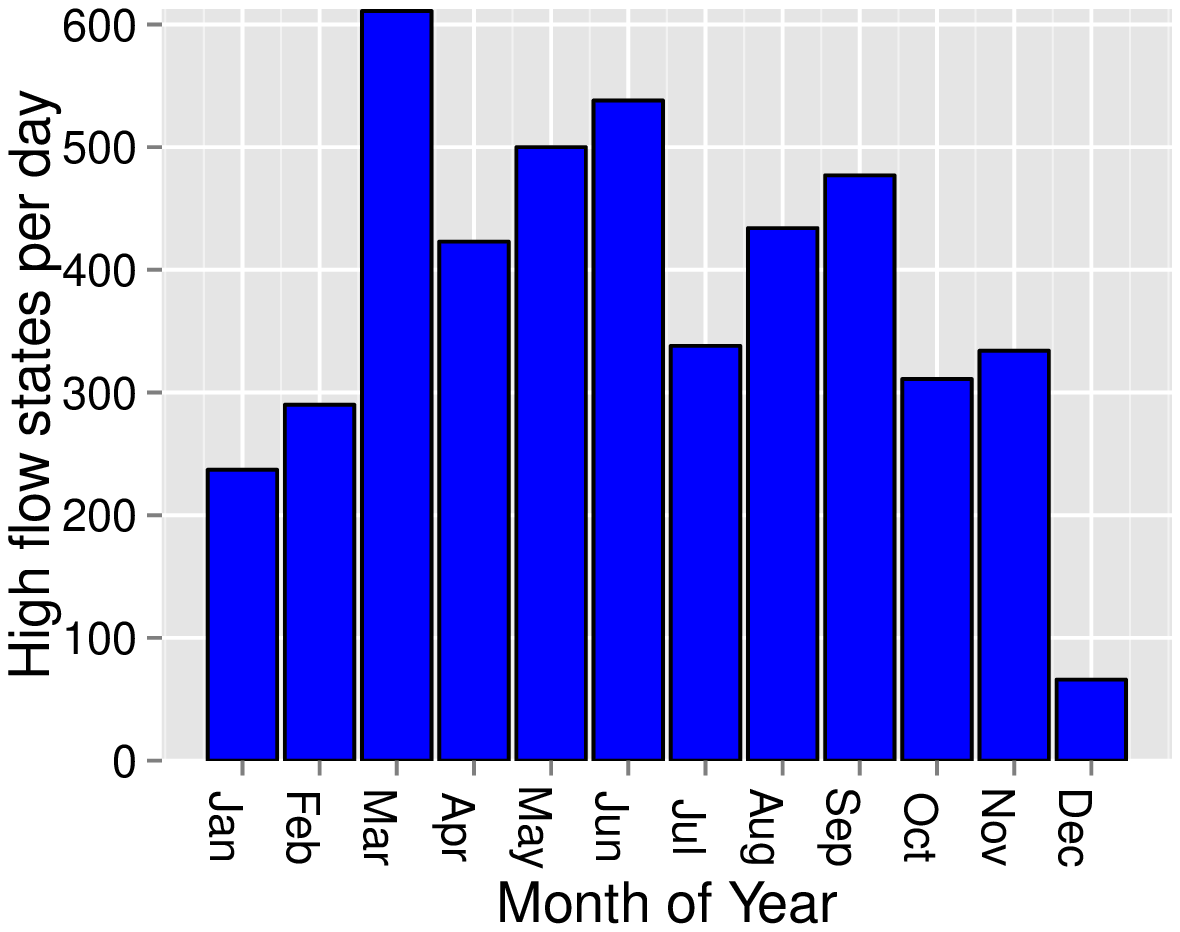}
		}
		\subfigure[]{%
			\label{fig:day}
			\includegraphics[width=0.46\textwidth]{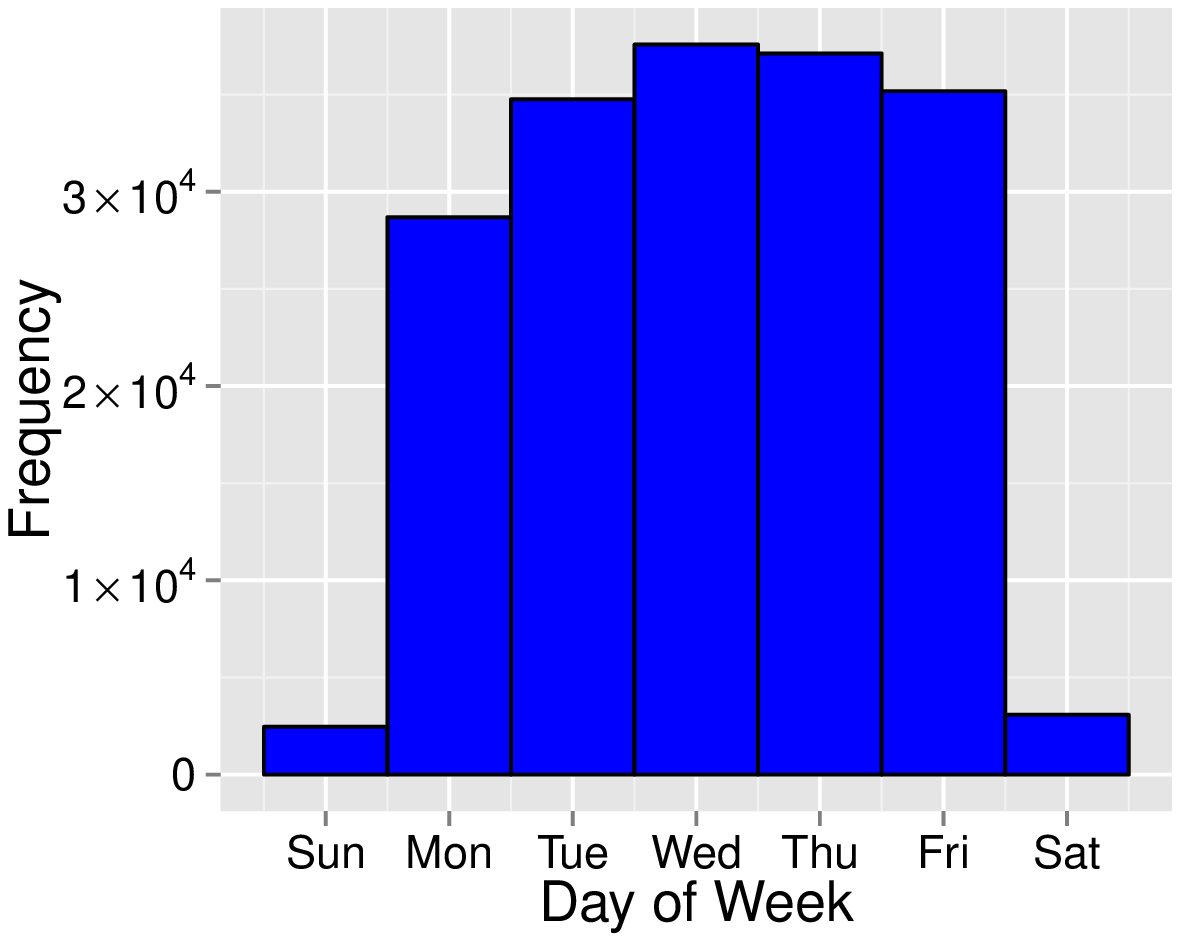}
		}\\
		\subfigure[]{%
			\label{fig:hour}
			\includegraphics[width=0.46\textwidth]{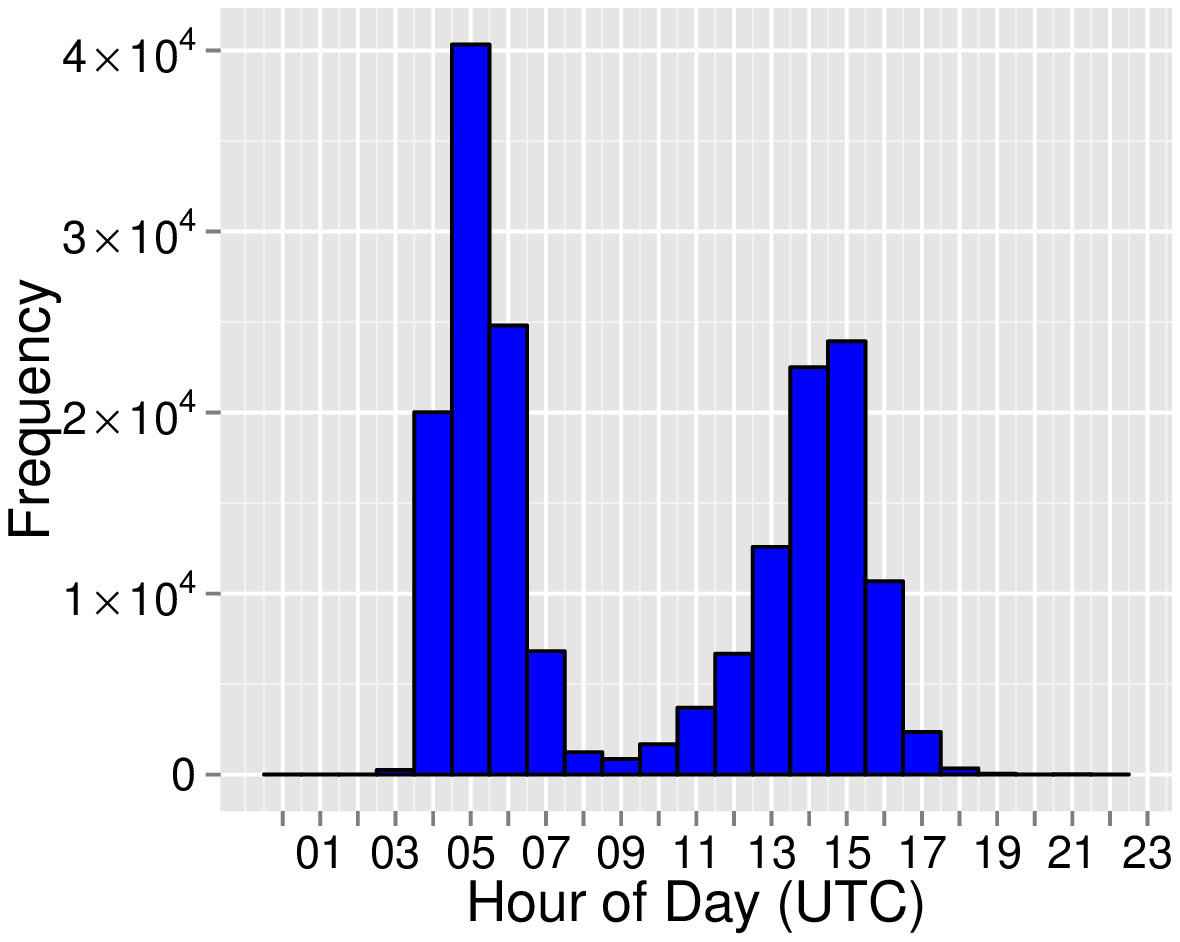}	 			
		}
		\subfigure[]{%
			\label{fig:lifetime}
			\includegraphics[width=0.46\textwidth]{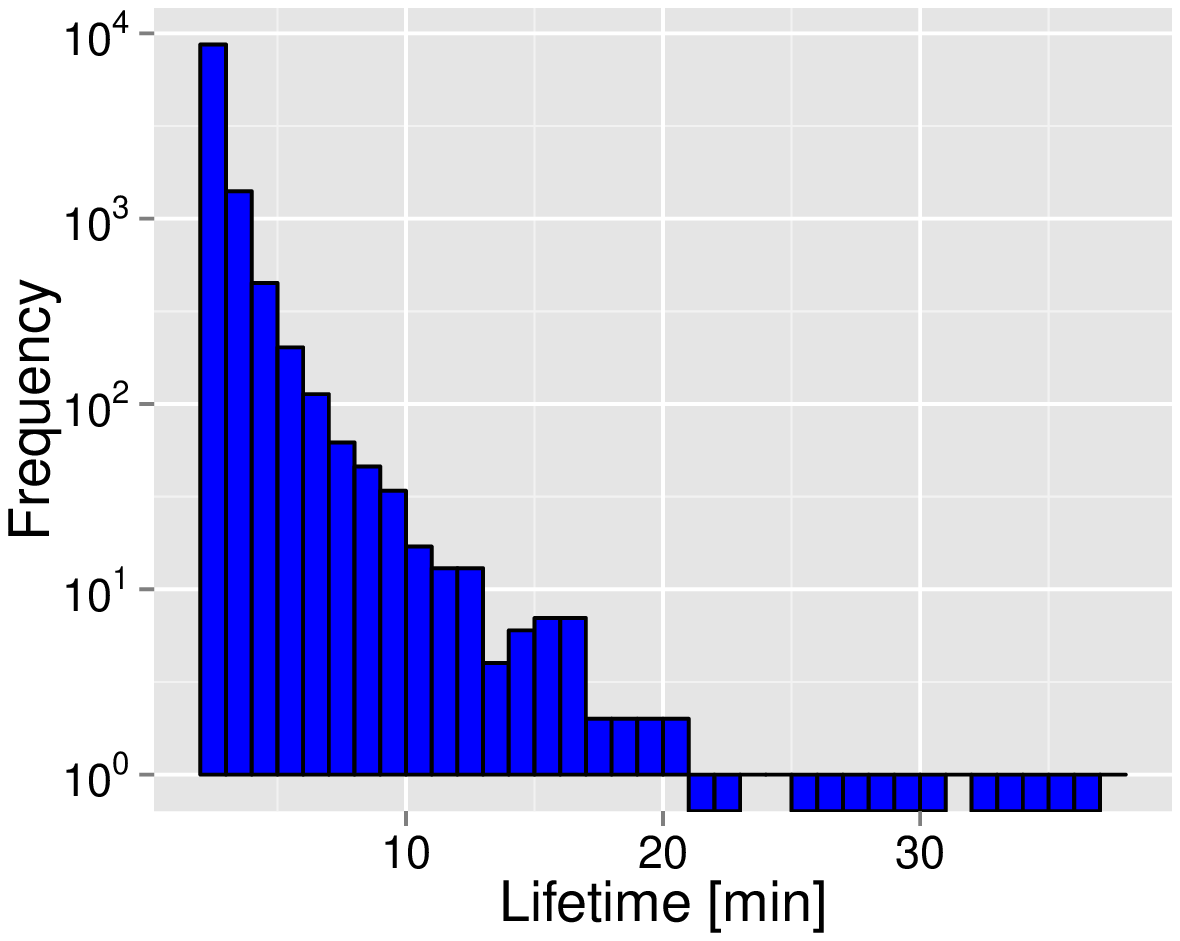}
		}		
		\caption{Temporal distribution of high-flow states depending 
					on \subref{fig:month} the month, \subref{fig:day} the weekday, 
					and \subref{fig:hour} the hour of day. 
					(The hour of day is given as UTC. The actual hour of day by adding 
					one or two hours---depending on daylight saving time.)
				 \subref{fig:lifetime} The frequency of successive measurements classifying as 
				 high-flow.}
		\label{fig:times}	
 	\end{figure}
\end{center}

The histograms given in figure~\ref{fig:times} show the temporal occurrence 
and the duration (i.e., lifetime) of high-flow states. 
If one considers that a high flow rate indicates a high traffic volume, the results of 
figures~\ref{fig:day} and \ref{fig:hour} are easy to understand. 
High-flow states occur on work days during peak-hours. 
At these times, there is a huge demand of commuters (i.e., many passenger cars) traveling 
to or from work. 

As the lifetime of a high-flow state we defined the number successive (1\,min)-intervals  
that were classified as ``high-flow state''. 
The resulting distribution of lifetimes is given in figure~\ref{fig:lifetime}. 
One can easily see that such states hardly last longer than a few minutes.
This  observation only confirms the long-known metastable character of traffic flow: 
An increased flow rate also increases the probability of a traffic 
breakdown~\cite{SchadschneiderChowdhuryNishinari2010,PersaudYagarBrownlee1998}. 
Especially at flow rates such as the ones considered in this article traffic flow is very unstable and long-lasting high-flow states could not be expected.

\section{Conclusions}
The findings of our analysis can be summarized as follows: 
high-flow states make high demands on the traffic conditions.
\begin{itemize}
        \item As a large number of vehicles is required for high-flow states, 
        such states are usually observed during the morning and evening peak hour
        on workdays (from Monday to Friday).
        \item The requirement of good road conditions is reflected by the fact that
        high-flow states are more likely to be observed in the summer.
        \item The share of trucks must be close to zero for high flows to occur as high average velocities are required.
        The lower the share of trucks, the higher is the average velocity.
        \item The lifetime of high-flow states is typically limited to intervals of 
        a few minutes length because traffic flow tends to be very unstable on this regime.
\end{itemize}
\bibliographystyle{iopart-num}
\bibliography{references.bib}


\printindex
\end{document}